\documentclass[final]{aipproc}
\layoutstyle{6x9}

\usepackage{amsfonts}
\usepackage{amsmath}
\usepackage{amssymb}

\newcommand{\be}{\begin{equation}}
\newcommand{\ee}{\end{equation}}
\newcommand{\ba}{\begin{eqnarray}}
\newcommand{\ea}{\end{eqnarray}}

\begin{document}

\title{$\eta/s$ is critical \\(at phase transitions)}


\author{Antonio Dobado, Felipe J. Llanes-Estrada and Juan M. Torres-Rincon}{address={Depto. F\'isica Te\'orica I, Universidad Complutense de Madrid,28040 Madrid, Spain}
}

\begin{abstract}
The viscosity over entropy density ratio, or KSS number, can help isolate the critical point in the hadron phase-diagram in Relativistic Heavy Ion Collisions. We argue that this quantity does have a minimum at a phase transition or crossover. Although indications from conventional non-relativistic gases point out to even a divergence in $\eta/s$ when the phase-transition is first-order,  since the critical exponent is rather low, this will be more difficult to ascertain in RHIC or FAIR. The experimental data are more likely to reveal a discontinuity for a first order phase transition or a smooth minimum at a crossover.
\end{abstract}

\maketitle

The quantity $\eta/s$ has become of interest after the realization that it might accept a universal lower bound $\eta/s \ge \frac{1}{4\pi}$ \cite{Kovtun:2003wp}, based on the AdS/CFT conjecture. That such a bound should exist was anticipated on the basis of the Heisenberg uncertainty principle and dimensional arguments \cite{Danielewicz:1984ww}, somewhat akin to Landauer's theory for conductivity. 
Empirically,  some examples that support this bound are Helium at the $\lambda$ point, where
$4\pi\eta/s$ is about 9, and for water in normal conditions, $4\pi\eta/s \simeq 380$.
It has been much discussed that the quark and gluon plasma is nearly a perfect fluid saturating the bound, with $\eta/s \sim 0.1$. In our work the low-temperature meson gas does satisfy this bound well thanks to unitarity \cite{Dobado:2006hw}.

We have recently shown \cite{Dobado:2008vt}, in agreement with  earlier authors \cite{Csernai:2006zz,Chen:2007xe,Lacey:2006bc},  that $\eta/s$, or KSS number (after the work of Kovtun, Son and Starinets
\cite{Kovtun:2003wp}), takes a minimum value near a phase transition.  This claim is supported by strong empirical evidence in conventional laboratory fluids, but also holds in several theoretical model results. This means that experimental extraction of $\eta/s$ in Heavy Ion Collisions can be a diagnostic to locate the phase transition to a quark and gluon plasma/liquid.

In these proceedings we also point out that, since the QCD phase transition is likely in the same dynamic universality class than the well-studied binary liquid mixture   or the gas-liquid transitions  \cite{Son:2004iv}, the critical behavior of the viscosity at the phase transition will be of power-law form,
\be
\frac{\eta}{s} = \left( \frac{\eta}{s}\right)_0 \ \left| \frac{T-T_c}{T_c}\right|^{-x_\eta},
\ee
where $(\eta/s)_0$ is the so-called critical amplitude and $x_{\eta}$ the critical exponent for the shear viscosity. This exponent has been measured \cite{soutocaride,sengers85,berg_moldover} in several nonrelativistic systems giving a similar result in all of them,  demonstrating the existence of an actual dynamic universality class. The critical exponent turns out to be very small, namely $x_{\eta} \simeq 0.041-0.042$.

Due to this tiny value we doubt this divergence will be found at FAIR, where $\eta/s$ can in principle be derived from experiment \cite{Gavin}.
However, the steep slopes left and right of the critical temperature mean that, in all likelihood, a discontinuity in $\eta/s$ will mark the phase transition.

Should a smooth behavior be found, this would be indicative of a crossover between hadron and quark phases. The smallest chemical potential for which a discontinuity be seen, is an upper bound to $\mu_c$ for the QCD critical endpoint.  

\begin{figure}[h]
\includegraphics[height=.245\textheight]{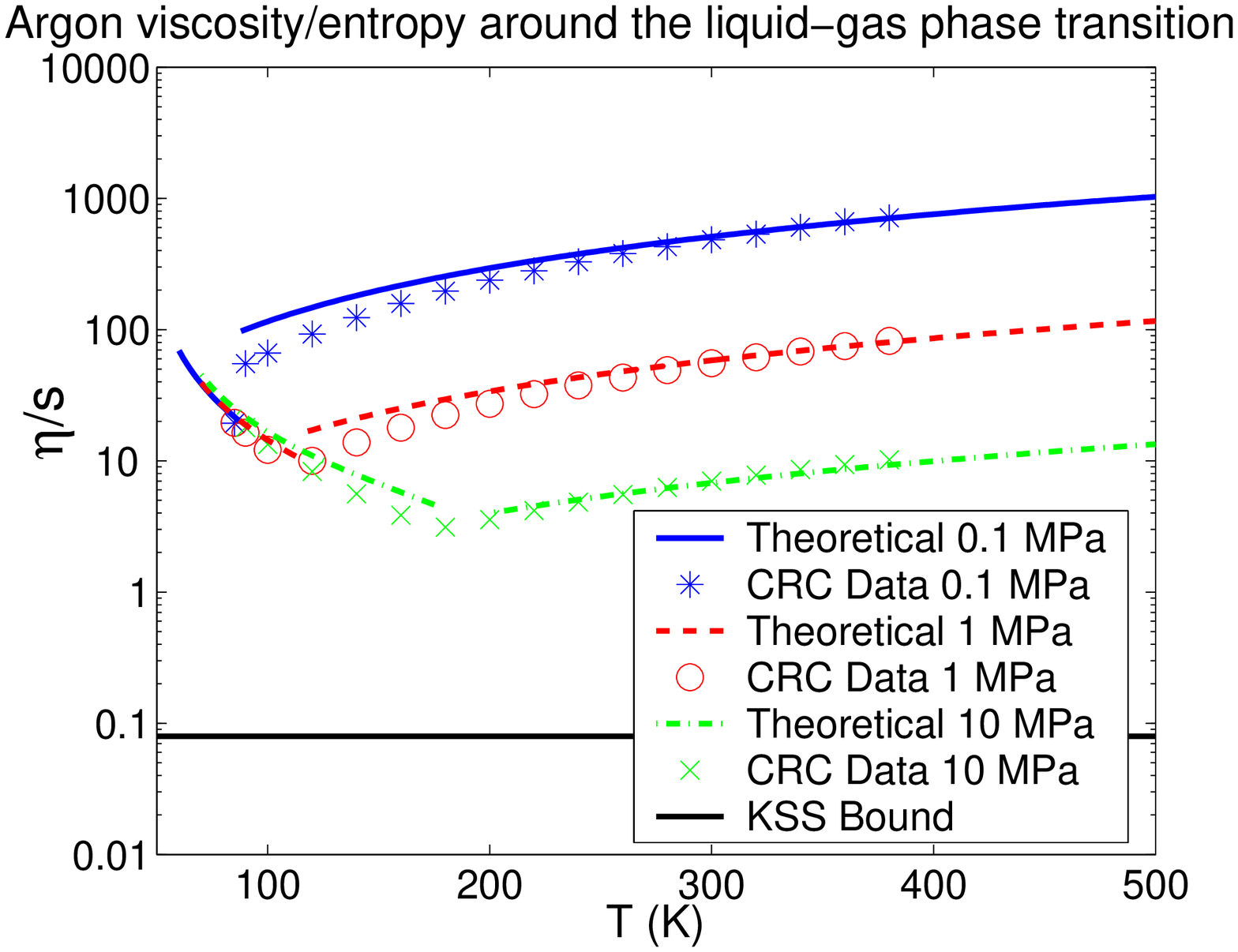}
\includegraphics[height=.245\textheight]{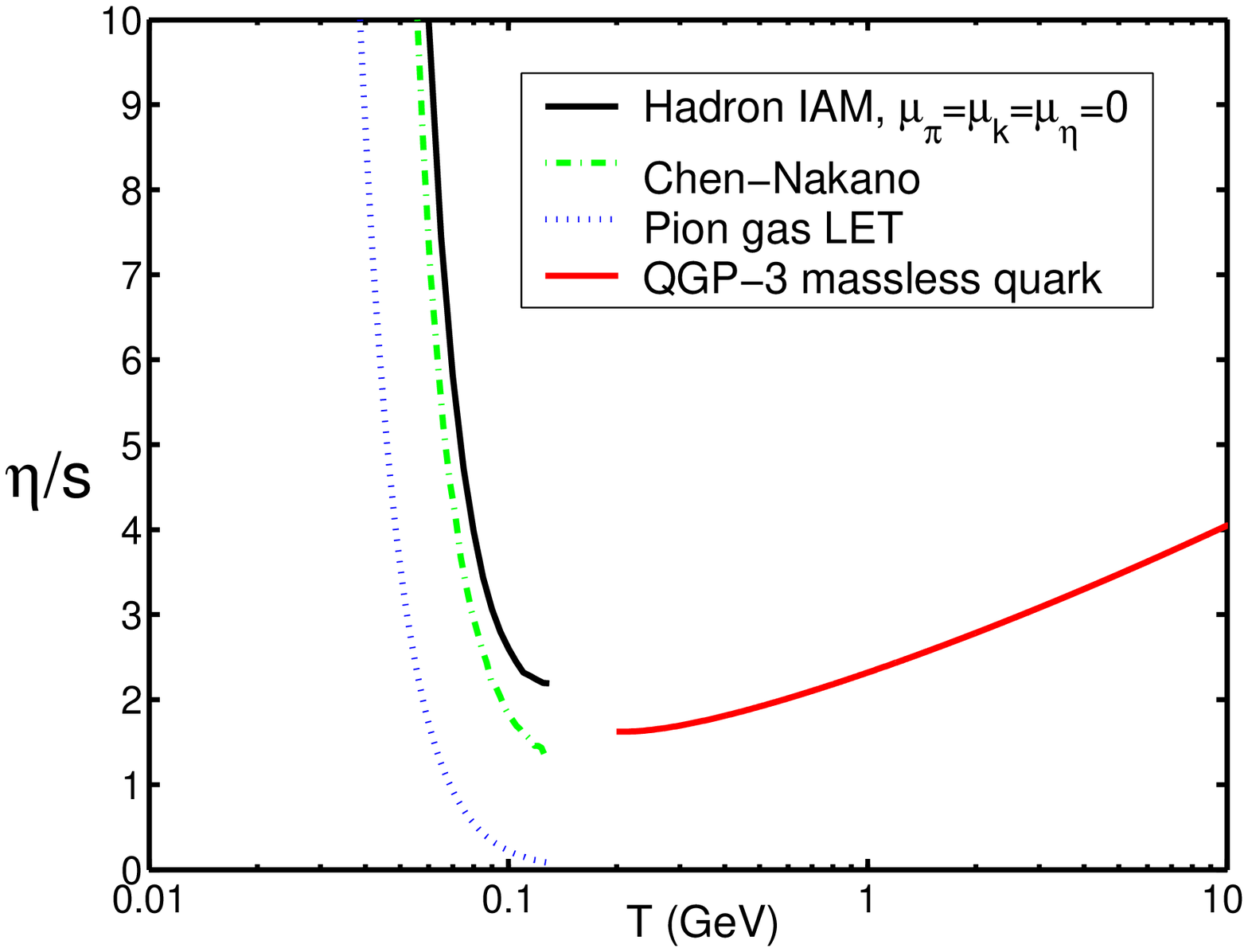}
\caption{$\eta/s$ (a pure number in natural units). Left: atomic Argon in the liquid and gas phases near the phase transition. 
Lines correspond to a theoretical calculation described in \cite{Dobado:2008vt}, the symbols are the experimental values from \cite{CRC}. 
Right: we improve the hadron-side (low $T$) estimate  of 
\cite{Csernai:2006zz} that showed the jump in the $\eta/s$ ratio in the 
transition from the hadron gas to the quark-gluon plasma, substituting 
the Low-Energy-Theorem of those authors (first order chiral perturbation 
theory) by the Inverse Amplitude Method, that agrees with Chiral 
Perturbation Theory at NLO, and satisfies elastic unitarity.
\label{arvoe}}
\end{figure}

This concept is illustrated in Figure \ref{arvoe}, 
where we show in the left panel the behavior of $\eta/s$ across the Argon gas-liquid phase transition. At large pressure, where a crossover links the hadron and possible quark phases, $\eta/s$ is seen to be continuous but to have a clear minimum. At smaller pressures a discontinuity builds up marking the first-order phase transition (note the entropy presents a  discontinuity and the viscosity a mild divergence, the ratio being discontinuous at low resolution). Argon is chosen because it is very close to a hard-sphere gas and amenable to theoretical treatment \cite{Dobado:2008vt}.
Note that $\eta/s$ is quite independent of the pressure in the liquid phase, and that the theoretical curves calculated from the liquid side and gas side do get closer together with increasing pressure, suggesting as the data that  indeed, $\eta/s$ will be continuous in the crossover regime.

We come now to the hadron phase transition. Our group \cite{Dobado:2003wr} has invested significant time in studying  transport coefficients in a meson gas with chiral perturbation theory, the Inverse Amplitude Method and both the Green-Kubo  and Boltzmann equation approaches. 
In the right panel we present a state of the art calculation of $\eta/s$ in the hadron gas at zero baryon chemical potential (where we expect a smooth crossover). In our approximation, the hadron gas is composed of $\pi$, $K$ and $\eta$ mesons interacting with relativistic kinematics and respecting unitarity in the collision process. Quantum effects have also been  incorporated in the transport equation.

Comparing with \cite{Csernai:2006zz} 
 we confirm the result of those authors, 
although the actual numerical value of $\eta/s$ is quite different (as 
should be expected from their calculation reaching temperatures $T\simeq
150$ MeV but with only the first order interaction). 
Comparing next to the quark-gluon plasma estimate on the right of the figure, we see that continuity in $\eta/s$ is conceivable with our calculation. We do not expect a strong discontinuity to appear at zero baryon chemical potential; however a minimum is likely because $\eta/s$ is known to grow for both $T\to 0$ (based on chiral perturbation theory) and $T\to \infty$ (based on a perturbative QGP estimate).

In conclusion, we have good reasons to believe that $\eta/s$ has a minimum (if the hadron to quark-gluon plasma/liquid is a crossover, as expected at low baryon density) or a divergence (if a first-order transition appears, typically at larger baryon densities), that, with finite experimental resolution, and given that the critical exponent will be small (0.04 for conventional fluids in the same universality class), will appear as a discontinuity. We look forward to future experiments addressing this interesting observable.


\begin{theacknowledgments}
  We thank the organizers of the \emph{XIII Mexican School of Particles and Fields} for  this 
very successful meeting. Work supported by grants BSCH-PR34/07-15875, FPA 2004 02602, FPA 2005-02327,
and Acci\'on Integrada Hispano-Portuguesa HP2006-0018.
\end{theacknowledgments}


\end{document}